# A hybrid material-point spheropolygon-element method for solid and granular material interaction

YUPENG JIANG*[1,2], MINCHEN LI[2], CHENFANFU JIANG[2], FERNANDO ALONSO-MARROQUIN[1]

1. School of Civil Engineering, The University of Sydney, Sydney, Australia
2. Department of Computer and Information Science, The University of Pennsylvania, Philadelphia, United States

**Abstract:** Capturing the interaction between objects that have an extreme difference in Young's modulus or geometrical scale is a highly challenging topic for numerical simulation. One of the fundamental questions is how to build an accurate multi-scale method with optimal computational efficiency. In this work, we develop a material-point-spheropolygon discrete element method (MPM-SDEM). Our approach fully couples the material point method (MPM) and the spheropolygon discrete element method (SDEM) through the exchange of contact force information. It combines the advantage of MPM for accurately simulating elastoplastic continuum materials and the high efficiency of DEM for calculating the Newtonian dynamics of discrete near-rigid objects. The MPM-SDEM framework is demonstrated with an explicit time integration scheme. Its accuracy and efficiency are further analysed against the analytical and experimental data. Results demonstrate this method could accurately capture the contact force and momentum exchange between materials while maintaining favourable computational stability and efficiency. Our framework exhibits great potential in the analysis of multi-scale, multi-physics phenomena.

1. INTRODUCTION

The numerical simulation of the multi-body system is crucial for understanding several key issues in geomechanics, such as the mechanical properties of the complex granular matrix[1-3], the interaction between the debris flow and solid structures[4-6]. It could also benefit the physics-based simulation in computer graphics (CG) to generate photorealistic visual effects for solid-fluid (or granular media) animation[7-8]. These systems commonly consist of individual bodies with disproportional sizes and various shapes. The fines-ballast granular structure typically exists at the railroad foundation[9] where the ratio between the volume of small (fines) and big particles (ballast) reaches a level of $1:10^5$. Components in the system could have distinctively different Young's modulus. For example, the crumb rubbers are artificially added to the railroad or highway foundation for preventing the breakage of surrounding granite particles and further improving the stability[10]. It is also common in CG production to animate intricate multi-body frictional contact between soft and rigid objects. The numerical methods for these issues often require an accurate simulation for the deformation of the individual bodies and contact forces among them. In other words, both the elastodynamic and the kinetic behaviours of each body need to be properly calculated. The Finite element method (FEM) is commonly applied, and proper treatment of the spatial discretization is essential for each component of the system[11, 12]. However, the accuracy and stability of FEM suffer from the potentially strong distortion of the meshes. Such a problem can be alleviated only by remeshing schemes[13] which, on the other hand, highly compromises the computational efficiency. The material points method (MPM) offers an attractive alternative approach. It discretises the computational domain with meshless particle and therefore avoids the difficulties encountered during large mesh deformation or topology changes. Because the deformation history of the material domain is stored and represented by material points, MPM utilizes a fixed Eulerian background grid that is not distorted during the simulation. MPM and its variants have been successfully applied for the study of continuum granular materials[14], crack propagation in snow[15], and computer animation[16].

As a numerical method based on continuum mechanics, either FEM or MPM requires a spatial discretization for every part of a multi-body system. Even for many engineering applications where only the kinematic information of near-rigid bodies is focused, the calculation of continuum partial differential equations still needs to be performed upon them[11]. The high stiffness of the material requires an extremely small time step interval for the stability of the simulation, while the convergence of iterative solvers in the implicit scheme is slow. Meanwhile, both methods use relatively complicated ways to handle the collision and contact force[17-19]. The computational efficiency could be largely improved if the near-rigid components are properly simplified so that only an accurate description of kinetic information is provided. The discrete element method (DEM) has high efficiency for simulating the Newtonian movements of rigid bodies. It treats near-rigid bodies, or particles, as perfectly rigid and defines an interaction zone that is coated outside each of them. The contact forces are calculated based on the depth, area or volume of the intersections among particle zones. This reasonable simplification ignores the deformation for the particles while accurately simulates the momentum and contact forces through contact laws. No spatial discretization or mesh generation is involved.

The coupling between DEM and other numerical methods (FEM, MPM) potentially provides an optimized balance between computational time and accuracy. Several methods have been systematically developed. For example, the DEM-FEM approach is conducted for hierarchical multiscale modelling of granular media and the interaction between the tire tread and granular terrain[21]. The coupling between DEM and Lattice Boltzmann method (LBM) is applied to study the interaction between water and soil particles in hydrological problems[22]. In these studies, discrete elements are simplified to circular/spherical shapes. Results of these simulations well agree with experimental data and analytical solutions. But still, geometrical parameters impose a strong influence on the movement of a near-rigid body. This factor has been further included in the numerical tools like DEM-CFD coupling between fluid and non-spherical particulate system[23].

One of the recent research for coupling of MPM-DEM was conducted by Liu et al[24]. The authors modelled a 2D sand pile collapsing and impacting three rectangular wooden blocks. The granular flow is simulated with MPM. A shrunken-point DEM is used to calculate the movement of blocks. Nine material points are attached at the corners, centre of the edges and the geometrical centre of each block. The contact between DEM blocks and the MPM flow are detected through the mutual background grid node. Contact forces are calculated at the mutual projection gird based on the momentum information and applied to geometrical nodes on the DEM parts as body forces. The numerical results show a reasonable agreement with the experimental data. This method could further help the damage analysis of buildings under the impact of debris flow. An MPM-DEM[25] hybrid method also developed to exploits the dual strengths of discrete and continuum treatments. However, this method is mainly focused on the improvement of the efficiency of DEM for simulating the granular flow. Discrete elements and material points are replacing each other under certain criterion rather than coexisting and interacting.

Liu et al's method[24] unifies the coupling procedures under the computational frame of MPM. It inevitably inherits an issue of MPM for handling the contacts, which is the strong dependency of the background grid system. First, the collision handling in MPM is calculated based on the proportions of momentum that different objects mutually contribute to grids. Contact detection and the calculation of force are fundamentally affected by the resolution and the structure of the grid system. Secondly, the contact algorithm needs to separately calculate the momentum for each object. The simulation will be computationally expensive if it involves a large number of discrete bodies. Meanwhile, the influence of shape is also omitted by simplifying the DEM particles with few material points. For many cases, such simplification is non-trivial and not rigorously discussed. The accuracy of the angular momentum of DEM particles is highly compromised since the forces are only applied at the geometrical nodes. These problems limit the performance of the coupling method and diminish the advantage of using DEM as a rigid body simulator. Therefore, we believe it is necessary to develop an advanced numerical method that could mitigate these issues and provide a more general way for the coupling between DEM and MPM.

In this paper, we develop a different coupling method between the DEM and MPM. The rigid body is represented by spheropolygon discrete element method (SDEM)[26]. It was developed for simulating the movement of irregular discrete particles. Comparing with other DEM-based methods, the SDEM could efficiently handle the contact forces among irregular particles and preserve the conservation of mechanical energy. Due to these merits, SDEM has been coupled with the boundary element method (BEM) for simulating sub-particle stress and particle breakage[27]. The interactions between the fluid and irregular rigid bodies are also studied using the LBM-SDEM[28]. In this work, the MPM is used for simulating the deformable part of the multi-body system. Two different constitutive models, linear elasticity and elastoplastic Drucker-Prager model are used to study the elastic-rigid and granular-rigid coupling respectively. The contact between SDEM and MPM are detected with the Euclidean distance instead of the existence of mutual projection grids. Coupling forces are directly calculated with well-established DEM contact model. For rigid particles, forces are applied at the exact contact position. In summary, our method unifies the coupling procedure under the computational frame of DEM. It significantly reduces the coupling dependency to the MPM background grid. The contact detection and force calculation only happen at the boundary of the rigid particles, which are more efficient than that of the pure MPM. The influence of particle shape can be better preserved and no longer needs to be simplified with material points.

The paper is organized as follows. The computational methodology for MPM and SDEM is introduced in Section 2 and 3 respectively. The coupling method is presented in Section 4. A serial of numerical tests is conducted in Section 5. Results are rigorously discussed with analytical solutions and experimental data as validations of the MPM-SDEM. Potential applications of this method are provided in Section 6. General conclusions are presented in Section 7.

## 2 MATERIAL POINT METHOD FOR ELASTIC AND GRANULAR MATERIAL

The material point method is a hybrid scheme utilizing both Lagrangian particles and Eulerian grids[29, 30]. It follows the governing equations of the continuum mechanics and its discretization is derived from the Galerkin weak form of momentum conservation, similarly to the finite element method. However, unlike the FEM, which discretises the computational area into piecewise subdomains on a mesh, the MPM uses the particle-wise material regions to represent the continuum. Lagrangian variables such as mass, momentum, and position are carried by the material points. The embedding relationship between Eulerian grids and material points is commonly defined by the nodal shape functions with $H^2$ regularity. At each time step, Lagrangian variables carried by material points need to be first transferred to the corresponding grid nodes. The equation of motion is solved at the grid nodes while volume integrals are approximated through particle quadrature, the velocity is updated accordingly and then interpolated back to material points for their advection and strain updates. Eulerian grids are restored to a standard Cartesian configuration after each time step, only the values and derivatives of the nodal shape functions are constantly recalculated at the beginning of the next time step.

### 2.1 Governing equations and discretization for MPM

The Lagrangian kinematic description of a continuum body needs to satisfy a group of partial differential equations (PDEs), including conservation of mass, momentum, and energy. These PDEs are known as governing equations which, combined with the material constitutive model and boundary conditions, determine the behaviour of the material. The conservation of mass is inherently satisfied in MPM since the material points in this study are assigned with constant mass values. The conservation of energy is guaranteed because the simulation assumes an isothermal setting which does not involve the exchange of heat. Therefore, the dynamic state of the material can be obtained by solving the conservation of moment[19,24,31]:

$$\sigma_{ij,j} + \rho b_i = \rho \ddot{u}_i, \tag{1}$$

where $\rho$ is the density of the material; $u_i$ denotes the displacement, the dots are the notation for the order of time derivative; $\sigma_{ij}$ is the Cauchy stress tensor, the subscript denotes the components and the deviator of the tensor; $b_i$ is the body force term. The PDEs follow Einstein notation. Equation 1 can be solved in the domain $\Omega$ through its weak form:

$$\int_\Omega u_i^*(\sigma_{ij,j} + \rho b_i - \rho \ddot{u}_i)\mathrm{d}\Omega = 0, \tag{2}$$

and boundary conditions:

$$\mathbf{t}(\mathbf{x})\big|_{\Gamma_t} = \mathbf{t}_0, \qquad \mathbf{u}(\mathbf{x})\big|_{\Gamma_u} = \mathbf{u}_0, \tag{3}$$

where the computational domain is denoted as $\Omega$; $u^*$ represents the virtual displacement which equals to zero on the boundary section $\Gamma_u$; the value of traction $\bar{t}_i$ is known at the boundary $\Gamma_t$. Combining the boundary conditions Equation 3 can be further written as:

$$\int_\Omega \rho \ddot{u}_i u_i^* d\Omega + \int_\Omega \sigma_{ij,j} u_i^* d\Omega - \int_\Omega \rho b_i u_i^* d\Omega - \int_{\partial\Omega} \bar{t}_i u_i^* d\Gamma. \qquad (4)$$

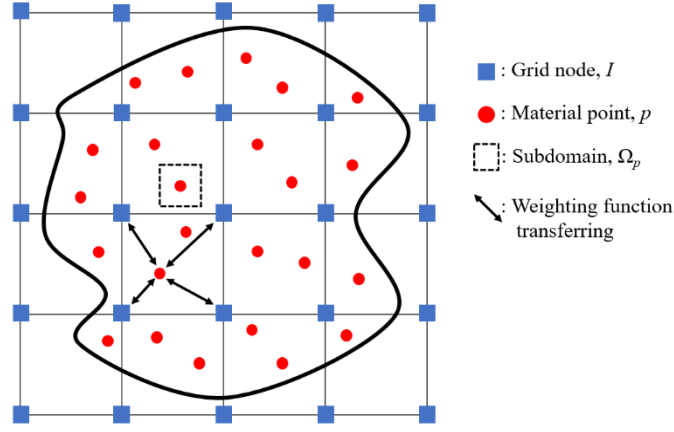

**FIGURE 1** Discretization schemes for the material point method

As shown in Figure 1, the domain is discretised by the material points (red dots), here also called "Lagrangian Points". the information of deformation gradient, mass and momentum are carried by these Lagrangian points. Eulerian background grid nodes (blue squares) are defined as background scratchpad. At each time step, the variables are first interpolated to grid nodes using multi-dimensional shape functions. The information is then updated at grid nodes and transferred back to the material points for the next time step. The perspective of generalized interpolation material point method (GIMP) is adopted for the discretization[32] process of the governing equation Eq. (4). Each material point occupies a partition $\Omega_p$ in the entity $\Omega$:

$$V_p = \int_{\Omega_p \cap \Omega} \chi_p(\mathbf{x}) d\Omega, \qquad (5)$$

where $V_p$ is the initial volume and $\chi_p(\mathbf{x})$ is the characteristic function, the subscript $p$ denotes the value on the material point. The mass of the material $m_p$ can be written as:

$$m_p = \int_{\Omega_p \cap \Omega} \rho_p(\mathbf{x}) \chi_p(\mathbf{x}) d\Omega, \qquad (6)$$

where $\rho_p = m_p/V_p$ is the density of the material. There are mainly two different forms of $\chi_p(\mathbf{x})$. One is using the total discrete approach (DMPM):

$$\chi_p(\mathbf{x}) = \delta(\mathbf{x} - \mathbf{x}_p) V_p, \qquad (7)$$

where $\mathbf{x}_p$ is the spatial coordinate of the material point and $\delta$ is the Dirac delta function. The mass only exits at the discrete position over the entire computational domain. The other method is to use the GIMP. where the form of $\chi_p(\mathbf{x})$ for each material point to be a continuous function (constant, linear or even a higher-order) over the $\Omega_p$. A given physical variable $k$ in the computational domain can be approximated by the value $k_p$ carried by the relevant material points and their characteristic functions:

$$k(\mathbf{x}) = \sum_p k_p \chi_p(\mathbf{x}), \qquad (8)$$

Equation (4) can be converted from the continuous integration form into a summation of material

points:

$$\sum_p \int_{\Omega_p \cap \Omega} \frac{\dot{p}_{ip}}{V_p} \chi_p u^*_{ip} d\Omega = -\sum_p \int_{\Omega_p \cap \Omega} \sigma_{ijp} \chi_p u^*_{ip,j} d\Omega$$

$$+ \sum_p \int_{\Omega_p \cap \Omega} \frac{m_p}{V_p} \chi_p b_{ip} u^*_{ip} d\Omega + \int_{\partial \Omega} \bar{t}_{ip} u^*_{ip} d\Gamma, \qquad (9)$$

The behaviours of the continuous material are now defined by the physical variables carried by material points. Equation 8 needs to be solved on the background grid nodes. The relation of virtual displacement between the grid node and material points is written as:

$$\mathbf{u}^*_p = \sum_I N_{Ip}(\mathbf{x}_p)\mathbf{u}^*_I, \qquad \mathbf{u}^*_{p,j} = \sum_I N_{Ip,j}(\mathbf{x}_p)\mathbf{u}^*_I. \qquad (10)$$

The subscript $I$ is the indexes that denote a value is on the grid node $I$. Substituting Equation 10 into Equation 9 to eliminate the virtual displacement and the equation of motion can be rewritten as:

$$\dot{\mathbf{p}}_I = \mathbf{f}^{int}_I + \mathbf{f}^{ext}_I, \quad \mathbf{x}_I \notin \Gamma_u \qquad (11)$$

where $\mathbf{p}_I$ is the momentum for the grid node; $\mathbf{f}^{int}_I$ and $\mathbf{f}^{ext}_I$ represent the internal and external force applied on the grid node respectively:

$$\dot{p}_{iI} = \sum_p \dot{p}_{ip} S_{Ip} \qquad (12)$$

$$f^{int}_{iI} = -\sum_p \sigma_{ijp} S_{Ip,j} V_p \qquad (13)$$

$$f^{ext}_{iI} = \sum_p m_p b_{ip} S_{Ip} + \int_{\partial \Omega} \bar{t}_i N_I d\Gamma \qquad (14)$$

The interpolation of variables between the material points and the grid nodes is calculated using the weighting function $S_{Ip}(\mathbf{x})$:

$$S_{Ip}(\mathbf{x}) = \frac{1}{V_p} \int_{\Omega_p \cap \Omega} \chi_p(\mathbf{x}) N_{Ip}(\mathbf{x}) d\Omega \qquad (15)$$

$$S_{Ip,j}(\mathbf{x}_p) = \frac{1}{V_p} \int_{\Omega_p \cap \Omega} \chi_p(\mathbf{x}) N_{Ip,j}(\mathbf{x}) d\Omega \qquad (16)$$

This interpolation is often referred as *P-G* (particle to grid) transferring and the reverse process is *G-P* (grid to particle) transferring. The specific forms of $S_{Ip}(\mathbf{x})$ depends on the choice of interpolation method (DMPM or GIMP) and shape function, which is provided in the Appendix. The interpolation function in GIMP has $C^1$ continuity even if the shape function is only a linear function with $C^0$ continuity. Generally, GIMP is more stable for spatial discretization, it produces less computational noises when the material points moving from one grid cell to another (cell-crossing noise). However, DMPM with quadratic or cubic B-spline weighting functions also has its advantage for being able to use a noise-free and angular momentum conserving transferring scheme called affine-particle-in-cell method (APIC)[33]. Note that as pointed out by Gao et al[7]., that GIMP and quadratic-B-Spline-DMPM are equivalent when particle domain is chosen to be a box with its width equal to the grid cell spacing.

## 2.2 Time integration scheme

The computational domain is time-variant, which indicates Equation 11 needs to be fulfilled at each time step. The central difference method is applied for the update of the grids' momentum:

$$\mathbf{p}_I^{n+1/2} = \mathbf{p}_I^{n-1/2} + \mathbf{f}_I^n \Delta t ,\qquad(17)$$

where superscript $n$ denotes the sequence of the time step and $\Delta t$ is a constant interval of the time increment at each step; $\mathbf{f}_I^n$ equals to $\mathbf{f}_I^{ext}+\mathbf{f}_I^{int}$ is the total nodal force. In this paper, we use the update-stress-first (USF)[34] format for the optimal stability and the conservation of energy of the simulation.

The algorithm to update the nodal and point variables is introduced in its computational sequences:
**a)** The nodal mass $m_I$, momentum $\mathbf{p}_I$, and velocity $\mathbf{v}_I$ are calculated through the interpolation of the corresponding material points and their weighting function $S_{Ip}$:

$$m_I^n = \sum_p m_p S_{Ip}^n ,\qquad(18)$$

$$\mathbf{p}_I^{n-1/2} = \sum_p m_p \mathbf{v}_p^{n-1/2} S_{Ip}^n ,\qquad(19)$$

$$\mathbf{v}_I^{n-1/2} = \mathbf{p}_I^{n-1/2} / m_I^n ,\qquad(20)$$

**b)** The material point's strain $\dot{\varepsilon}_{ijp}^{n-1/2}$ and spin $\dot{\psi}_{ijp}^{n-1/2}$ rate tensors are obtained with the nodal velocities[24]:

$$\dot{\varepsilon}_{ijp}^{n-1/2} = \frac{1}{2}\sum_I (S_{Ip,j}^n v_{iI}^{n-1/2} + S_{Ip,i}^n v_{jI}^{n-1/2}) ,\qquad(21)$$

$$\dot{\psi}_{ijp}^{n-1/2} = \frac{1}{2}\sum_I (S_{Ip,j}^n v_{iI}^{n-1/2} - S_{Ip,i}^n v_{jI}^{n-1/2}) ,\qquad(22)$$

where the value of $S_{Ip}$ is decided according to the method of spatial discretization and the relative position between the material point and grid node. The strain and stress are then calculated as:

$$\varepsilon_{ijp}^n = \varepsilon_{ijp}^{n-1} + \dot{\varepsilon}_{ijp}^{n-1/2} \Delta t ,\qquad(23)$$

$$\sigma_{ijp}^n = \sigma_{ijp}^{n-1} + \dot{\sigma}_{ijp}^{n-1/2} \Delta t .\qquad(24)$$

The stress rate is determined by:

$$\dot{\sigma}_{ij} = \sigma_{ij}^{\nabla} + \sigma_{ik}\psi_{jk} + \sigma_{jk}\psi_{ik} , \quad \sigma_{ij}^{\nabla} = C_{ijkl}\dot{\varepsilon}_{kl} ,\qquad(25)$$

where Jaumann stress rate $\sigma_{ij}^{\nabla}$ for linear elasticity is adopted to eliminate the influence of pure rotation to the Cauchy stress tensor. The specific form of stress rate $\dot{\sigma}_{ij}$ depends on the constitutive model.
**c)** The internal and external nodal forces are updated using the material point's stress, the body

force term $\mathbf{b}_p$, and the boundary traction $\mathbf{t}_p$ according to Equation 12 and 13

**d)** Equation 11 is solved at the grid nodes with the nodal forces and momentum to further update the velocity and position of material points. There are three methods for this specific G-P transferring procedure, particle-in-cell format (PIC)[29,35], fluid-Implicit-particle format (FLIP)[36] and the hybrid format. The PIC transferring update the material's velocity directly using the interpolation functions and corresponding grid values:

$$\mathbf{v}_p^{n+1/2} = \sum_I \mathbf{p}_I^{n+1/2} S_{Ip}^n / m_I^n . \qquad (26)$$

This transferring method has a strong numerical dissipation; the angular momentum and kinetic energy of material points decrease rapidly with the progress of iterations. This problem makes the PIC format unsuitable for simulating dynamic problems such as the landslide. Because the kinetic energy of granular flows is not accurate enough and hence make the results unreliable. However, this dissipation effect is not entirely undesirable. It efficiently decreases the vibration in quasistatic simulations, which generated by the initial configuration of the objects, and help the system to reach an equilibrium. The FLIP updates the velocity using the nodal force:

$$\mathbf{v}_p^{n+1/2} = \mathbf{v}_p^{n-1/2} + \Delta t \sum_I \mathbf{f}_I^n S_{Ip}^n / m_I^n . \qquad (27)$$

It greatly improves the conservation of angular momentum. But the exact conservation can only be achieved by using the 'full' mass matrix instead of lumped mass matrix, which is necessary for the numerical stability but impractical due to its potential singularity[33,37]. Meanwhile, FLIP format generates the so-called "ringing instability" because of noisy velocity modes in the null space of the transfer operator. It causes the numerical instability when simulating the granular material with strong dynamic behaviours. A better way to preserve the advantages of PIC and FLIP is to use a hybrid transferring format[38]:

$$\mathbf{v}_p^{n+1/2} = \alpha \sum_I \mathbf{p}_I^{n+1/2} S_{Ip}^n / m_I^n + (1-\alpha)(\mathbf{v}_p^{n-1/2} + \Delta t \sum_I \mathbf{f}_I^n S_{Ip}^n / m_I^n), \qquad (28)$$

where $\alpha$ is the coefficient that ranges between 0 to 1; $\alpha=0$ is the pure FLIP format and $\alpha=0$ gives the pure PIC format. The value of $\alpha$ is proportional to the magnitude of the dissipation effect and can be used for controlling the damping for the quasi-static simulation. The instability of FLIP format is also alleviated. However, the hybrid method still has an issue. No rigorous analysis to define the quantitative relation between the dissipation and the value of $\alpha$. The behaviour of the material is also sensitive to the variation of $\alpha$ and the numerical damping highly dependents on the time step size. Especially for the granular material, the specific value of the coefficient is rather an empirical setting.

Such a problem can be properly solved by using affine-particle-in-cell format (APIC) or extended-particle-in-cell format (XPIC)[39]. We adopt APIC in this paper. This innovative method is developed by Jiang and Schroeder[31]. It represents particle velocities as locally affine, which allows APIC to conserve linear and angular momentum across transfers. This transferring format effectively reduces the numerical dissipation; it also does not experience the velocity noise and instability in FLIP. It has been applied for the simulation of both the granular and hyper-elastic material and exhibits superior performance. The APIC still use the PIC format for the velocity and position update of material points at step **(d)**. However, APIC applies a different scheme for the *P-G* transferring procedure at step **(a)**, which is written as[33,40]:

$$\mathbf{D}_p^n = \sum_I N_{Ip}^n (\mathbf{x}_I^n - \mathbf{x}_p^n)(\mathbf{x}_I^n - \mathbf{x}_p^n)^T = \begin{cases} L^3\mathbf{I}/3 & \text{cubic} \\ L^2\mathbf{I}/4 & \text{quadartic} \end{cases}, \quad (29)$$

$$\mathbf{p}_I^{n-1/2} = \sum_p m_p N_{Ip}^n (\mathbf{v}_p^{n-1/2} + \mathbf{B}_p^n (\mathbf{D}_p^n)^{-1}(\mathbf{x}_I^n - \mathbf{x}_p^n)), \quad (30)$$

where $L$ is the grid spacing and $\mathbf{I}$ is the unit matrix; the additional matrix $\mathbf{D}_p$ serves as the inertia matrix for affine motion and $\mathbf{B}_p$ contains the angular momentum information and updates as:

$$\mathbf{B}_p^{n+1/2} = \sum_I S_{Ip} v_I^{n+1/2} (\mathbf{x}_I^n - \mathbf{x}_p^n)^T \quad (31)$$

In this paper, we use the GIMP with hybrid format for simulating the linear elastic material and DMPM with APIC format for the granular material. The quadratic kernel is adopted as the shape function.

### 2.3 Constitutive models

The physical properties of granular material have been a difficulty for the numerical simulation. Large deformation may happen due to the elastoplastic behaviour of granular material which causes trouble for the mesh-based method. The MPM, therefore, has a unique advantage for simulating both quasistatic deformation and granular flow without the restriction of mesh. The Drucker-Prager plasticity model(*D-P*) is employed to simulate the granular material. This model has been widely applied for the engineering application and MPM sand simulation. Although the Jaumann stress rate introduced earlier may not be exceedingly accurate for moderate deformation with deviatoric strains that more than 10 percent (e.g. granular flow)[41], it can still provide reasonable results and we choose to use it for a direct comparison to the results in the existing literature[24].

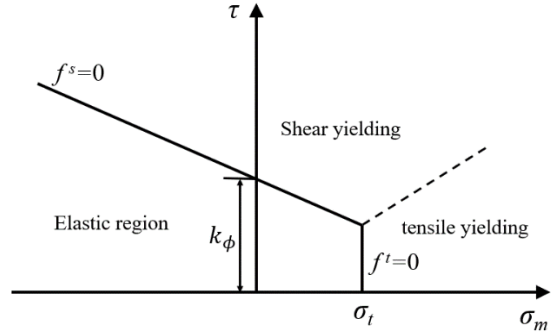

**FIGURE 2** Yield criteria of Drucker-Prager(*D-P*) model[24]

The yield criteria in *D-P* model is shown in Figure 2 and defined as:

$$f^s = \tau + q_\phi - k_\phi, \quad (31)$$

$$f^t = \sigma_m - \sigma^t, \quad (32)$$

where $f$ is the yielding surface, the superscripts $s$ and $t$ denote the shear and tensile yielding behaviour respectively; $\sigma^t$ is the tensile strength; $\tau$ is the equivalent shear stress and $\sigma_m$ is the spherical stress:

$$\tau = \sqrt{J_2}, J_2 = s_{ij}s_{ij}/2, \quad (33)$$

$$\sigma_m = I_1/3, I_1 = \sigma_{kk} \qquad (34)$$

where $J_2$ is the second invariant of the deviatoric stress tensor and $s_{ij}$ is the deviatoric stress components; $I_1$ denotes the first invariant of the stress tensor. The coefficients $q_\phi$ and $k_\phi$ are frictional coefficient and yield stress for shearing behaviour. They are calculated based on the friction angle $\phi$ and the cohesion term $c$:

$$q_\phi = \frac{3\tan\phi}{\sqrt{9+12\tan^2\phi}}, k_\phi = \frac{3c}{\sqrt{9+12\tan^2\phi}}. \qquad (35)$$

MPM uses the return mapping[24] method to detect whether the yielding conditions are fulfilled at each time step. The isotropic linear elasticity is used for solid material:

$$C_{ijkl} = K\delta_{ij}\delta_{kl} + G(\delta_{ik}\delta_{jl} + \delta_{il}\delta_{jk} - \frac{1}{3}\delta_{ij}\delta_{kl}), \qquad (36)$$

where **C** is the constitutive tensor; $K$ and $G$ are the bulk and shear module respectively.

## 3. SPHEROPOLYGON DISCRETE ELEMENT METHOD (SDEM)

The spheropolygon discrete element method[25] (SDEM) is selected as the DEM part of the hybrid algorithm due to its unique advantages. The geometrical irregularity of a particle is properly represented as a Minkowski sum of a polygon with a disk, and the multiple contacts between irregular particles are calculated based on distances between vertices and edges. These features make it computationally efficient[25]. The contact information between MPM and the SDEM can be easily applied to the correct position of SDEM instead of its geometrical modes. The point contact relation is also more realistic for contacts between rigid bodies and granular materials.

### 3.1 SDEM algorithm

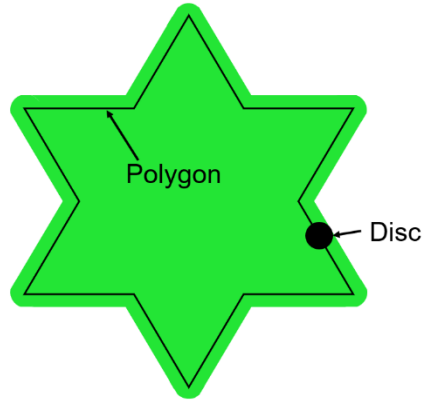

**FIGURE 3** Spheropolygon element obtained by sweeping a disk around a polygon.

A spheropolygon is the Minkowski sum of a polygon to represent the irregular shape of near-rigid object and a disk with radius $a$, which defines an elastic area to calculate the contact forces that generated among particles. Mathematically the Minkowski sum of two sets of points $P$ and $Q$ of a vector space is given by[26]:

$$P+Q = \{\mathbf{x}+\mathbf{y} | \mathbf{x}\in P, \mathbf{y}\in Q\}. \qquad (37)$$

The geometrical interpretation of this operation is equivalent to the sweeping of a disk around the profile of the polygon while maintaining its original orientation. For example, the DEM particle with a shape of hexagram in Figure 3 is properly approximated by a spheropolygon element with a few boundary lines and a disk sweeping its profile. The hexagram defines the element shape, and the disk is used for contact force calculation. SDEM has been proved as a more effective approach than using a cluster of small particles to approach this shape, especially when the particle shape is more irregular or complicated.

The contact force of the spheropolygon is defined through a vertex-edge contact relationship. Let us consider two spheropolygons $SP_i$ and $SP_j$ with their polygons $P_i$ and $P_j$ and the radii of the disks $a_i$ and $a_j$. Each polygon is defined by its own set of vertices $\{V\}$ and edges $\{E\}$. The overlapping length $\xi$ between each vertex-edge pair $(V, E)$ is written as:

$$\xi(V,E) = \langle a_i + a_j - d(V,E) \rangle, \tag{38}$$

where $d(V, E)$ is the Euclidean distance between the vertex $V$ and the edge $E$. The brace at the right side of the equation means the non-negative limit of $\xi$. Therefore, the force vector $\mathbf{F}$ applied on particle $i$ by particle $j$ is expressed as:

$$\mathbf{F}_{ij} = -\mathbf{F}_{ji} = \sum_{V_i E_j} \mathbf{F}(V_i, E_j) + \sum_{V_j E_i} \mathbf{F}(V_j, E_i), \tag{39}$$

and the torque $\boldsymbol{\tau}_{ij}$ of particle $i$ is:

$$\boldsymbol{\tau}_{ij} = \sum_{V_i E_j} (\mathbf{p}(V_i, E_j) - \mathbf{c}_i) \times \mathbf{F}(V_i, E_j) + \sum_{V_j E_i} (\mathbf{p}(V_j, E_i) - \mathbf{c}_i) \times \mathbf{F}(V_j, E_i), \tag{40}$$

where $\mathbf{c}_i$ is the centre of mass of particle $i$ and $\mathbf{p}$ is the point of contact, which is defined as the middle point of the overlap area between a vertex and an edge:

$$\mathbf{p}(V,E) = \mathbf{X} + (a_i - \frac{1}{2}\delta(V,E)) \frac{\mathbf{X}-\mathbf{Y}}{\|\mathbf{Y}-\mathbf{X}\|}, \tag{41}$$

where $\mathbf{X}$ is the position of the vertex $V$, and $\mathbf{Y}$ is its closest point on the edge $E$. The movement of the centre of mass $r_i$ and the orientation $\varphi_i$ of the particle are governed by the equations of motion:

$$m_i \ddot{\mathbf{c}} = \sum_j \mathbf{F}_{ij}, \quad I_i \ddot{\boldsymbol{\varphi}} = \sum_j \boldsymbol{\tau}_{ij}, \tag{42}$$

where $m_i$ and $I_i$ are the mass and moment of inertia of the particle; The linear elastic model is used throughout this paper. The force $\mathbf{F}$ can be written as:

$$\mathbf{F} = k_n \delta_n \mathbf{N} + k_t \delta_t \mathbf{T}, \tag{43}$$

where $k_n$ and $k_t$ are the normal and tangential stiffness, respectively; $\mathbf{N}$ and $\mathbf{T}$ are the normal and tangential unit vector that measured at the Edge of the contact; $\delta_n$ denotes the length of the overlap; $\delta_t$ is the tangential relative displacement between two particles that accounts for frictional forces. It is limited by the condition $k_t \delta_t \leq \mu k_n \delta_n$, where $\mu$ is the friction coefficient.

## 4. COUPLING of MPM AND SDEM

MPM and SDEM are fully coupled through the contact force information. The crucial part of the coupling is how to properly detect and calculate the contact force between the material points and

SDEM particles. Liu et al proposed a method which attaches material points to the DEM particles; the square particles are represented with 9 materials at its corners, edges' middle point and centre of mass. Both the contact detection and contact forces are then calculated using a pure MPM contact algorithm developed by Bardenhangen and Brackbill[14]. This coupling method unifies the contact handling under a pure MPM scheme. The fundamental reason for these problems rises from the simplification of the contact and the dependency of the grid system. The contact force cannot be accurately calculated if only a small amount of material points is involved. However, the advantage of efficiency largely decreases if the material points are densely attached to the DEM particle.

### 4.1 Contact handling between MPM and SDEM

In this paper, we proposed a different approach for the computation of contact handling, which unifies the contact detect and force calculation under the scheme of SDEM. The basic idea of this algorithm can be summarized as follows. The Verlet distance, which is the cut-off distance for the potential contact between two discrete elements, is applied to examine the contact between material points and SDEM particles. If a material point is within the Verlet distance of a spheropolygon particle it is treated as a small SDEM disc particle with a certain radius for contact detection. If the material points and spheropolygon is in contact, the contact force between the material point and the spheropolygon is calculated based on DEM contact force models. The calculated force is applied to the material points in a form of extra boundary force term.

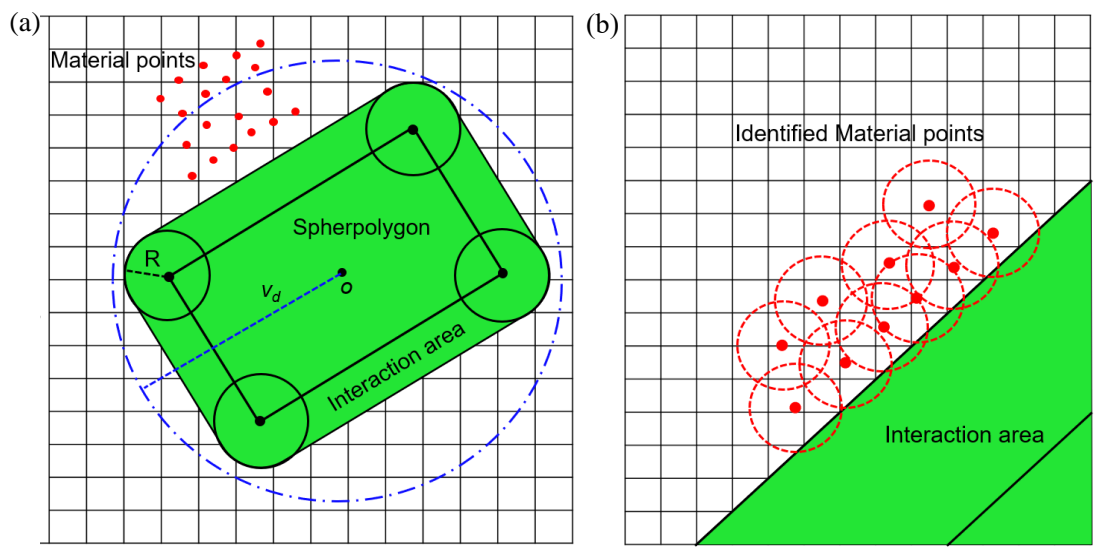

**FIGURE 4** The contact handling between the spheropolygon and material points (a) Spheropolygon is represented by the green zone and material points by red points, $V_d$ is the Verlet radius from the centre of mass $o$. (b) material points in potential contact with the spheropolygon are assigned with a small radius (red dash circles).

As indicated in Figure 4 (a) material points within the Verlet distance $V_d$ are transformed into a circular discrete element with its position $\mathbf{x}_p$ as the centre of mass. Here we name it *identified material points* (IMP). The interactions among the IMPs are still calculated under the MPM scheme despite the intersections of their radius may happen after the transformation. Figure 5 (a) shows that the magnitude of the contact force $\mathbf{f}^{SDEM}$ between an IMP and an SDEM particle can be then calculated based on the modification of Equation 39:

$$\xi(\mathbf{x}_p, E) = \langle a_i + r_p - d(\mathbf{x}_p, E) \rangle, \tag{44}$$

where $r_p$ is the contact radius assigned to an IMP; $a_i$ is the sphero-radius of the SDEM particle; $d$ denotes the Euclidean distance between the points $\mathbf{x}_p$ and the edge $E$ on the SDEM particle. The optimal value of $r_p$ is still an open question. Here we provide an estimated interval for $r_p$:

$$l_p \sqrt{1/n} < r_p < l_p, \tag{45}$$

where $l_p$ is the length of the background grid and $n$ is the average number of the material point in each grid. The value of $r_p$ must be smaller than the grid length to control the contact happens within or at the boundary of a cell. Meanwhile, it should be larger than the influential length that a material point representing in the computational domain.

If an IMP is in contact with multiple SDEM particles, as shown in Figure 5 (b), the particle-spheropolygon forces are summed together. The single contact of IMP is most likely to happen during the coupling since the radius assigned to the IMP is much smaller than the size of SDEM particles. But in extreme cases, such multiple contact relationship will not affect the stability or the efficiency of the coupling.

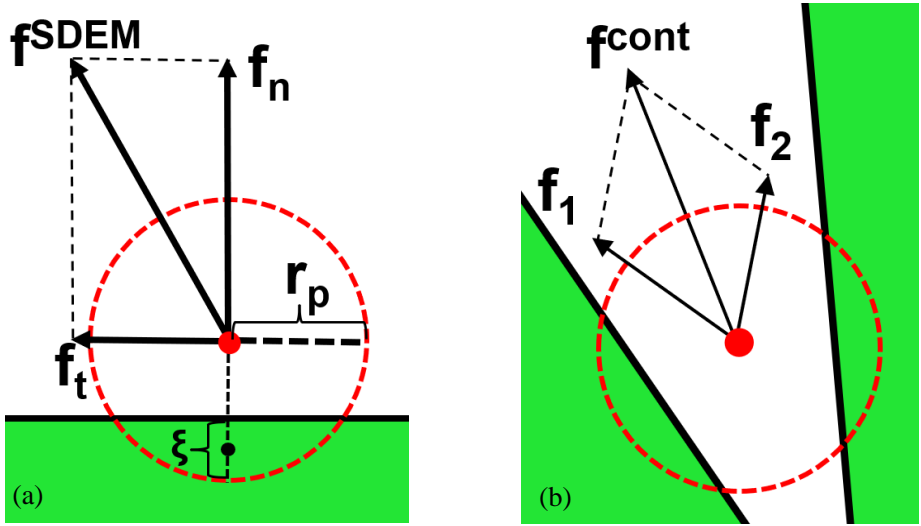

**FIGURE 5** The intersection and contact force calculation between an IMP and SDEM. (a) The calculation of the contact force between one IMP and SDEM particle;(b) the total forces exert on an IMP

The contact forces are applied at the corresponding material point as an external boundary force:

$$\mathbf{f}_p^{cont} = \sum \mathbf{f}_k^{SDEM}, \tag{46}$$

$$\mathbf{f}_I^{cont} = \sum_p \mathbf{f}_p^{cont} S_{Ip}(\mathbf{x}_p), \tag{47}$$

where the $\mathbf{f}^{SDEM}$ is the coupling contact force; $\mathbf{f}_p^{cont}$ and $\mathbf{f}_I^{cont}$ is the external force on the IMPs and corresponding nodes respectively. Therefore, the nodal equation of motion (Equation 18) is further modified as:

$$\dot{\mathbf{p}}_I = \mathbf{f}_I^{int} + \mathbf{f}_I^{ext} + \mathbf{f}_I^{cont}. \tag{48}$$

In this way, the influence of the contact force is taken into the calculation of MPM at step (**c**). And through this equation, it could further its influence on the movement of material points in the whole computational area. For the SDEM particle, the contact forces are applied at its centre of mass. The total force and torque exerted on the particle in Equation 39 and 40 are now modified as:

$$\mathbf{F}_{ij} = \sum_{V_i E_j} \mathbf{F}(V_i, E_j) + \sum_{V_j E_i} \mathbf{F}(V_j, E_i) + \sum_p \mathbf{f}^{SDEM}(V_i, \mathbf{x}_p) + \sum_p \mathbf{f}^{SDEM}(\mathbf{x}_p, E_i), \tag{49}$$

$$\boldsymbol{\tau}_{ij} = \sum_{V_i E_j} (\mathbf{p}(V_i, E_j) - \mathbf{c}_i) \times \mathbf{F}(V_i, E_j) + \sum_{V_j E_i} (\mathbf{p}(V_j, E_i) - \mathbf{c}_i) \times \mathbf{F}(V_j, E_i)$$
$$+ \sum_p (\mathbf{p}(V_i, \mathbf{x}_p) - \mathbf{c}_i) \times \mathbf{f}^{SDEM}(V_i, \mathbf{x}_p) + \sum_p (\mathbf{p}(\mathbf{x}_p, E_i) - \mathbf{c}_i) \times \mathbf{f}^{SDEM}(\mathbf{x}_p, E_i). \tag{50}$$

These equations equal to adding extra contact forces that produced by small circular particles to the large SDEM particle, so that the movement and rotation of the SDEM particles are also affected by the IMP that is in contact with it.

Since the algorithm is using an explicit form the critical time step $\Delta t_{min}$ needs to be determined for the stability of the coupling. This criterion[42] can be written as:

$$\Delta t_{min} = \min \begin{cases} \kappa_1 \dfrac{l_{min}}{c_{max}} \\ 2\pi \kappa_2 \sqrt{\dfrac{m_{min}}{k_n}} \end{cases} \tag{51}$$

where the $l_{min}$ is the minimum length of the background grid and $c_{max}$ is the maximum acoustic velocity of the material in MPM; $m_{min}$ is the minimum mass of the SDEM particle. The coefficient $\kappa_1$ and $\kappa_2$ are used to further guarantee the stability since the equation for critical time step is obtained based on the linear elasticity, which $\kappa_1=0.8$ and $\kappa_2=0.1$ are commonly used.

The contact between material points and the SDEM particle can be fully coupled together through the contact force. There are several advantages to using this scheme. Contact relations are detected with the Euclidian distance between the centre of mass of SDEM particle and the position of a material point. The contact list can be saved as a data structure and reused for the next time step if the displacement of the particles and material points is small. It is more efficient than the condition of mutual grid node where the velocity field on each object is constantly recalculated. Contact forces can be directly calculated with DEM contact models instead of using the grid momentum as an indirect approach. Positions for applying contact forces are not restricted by the grid node or the geometrical nodes of the rigid particles. Irregular shapes are considered, and the angular momentum of the rigid body is better preserved.

**4.2 Coupling method procedures.**

The program for MPM-SDEM is developed (MPM-SPOLY) using C++ language for further validations and applications. The coupling algorithm in a time step $\Delta t$ can be summarized as follows:

(i) **SDEM**: contact detection and force calculation
    Update the contact list of SDEM particles based on the Verlet distance.
    Update the contact list of SDEM particles and IMPs.
    Update Vertices {$V$} for each particle
    Update the Vertex-edge contact relations between particles within the contact list
    Calculate the contact force of SDEM particles and apply the gravity
    Calculate the $\mathbf{f}^{SDEM}$ between SDEM particles and IMPs

(ii) **MPM**: variables transfer between nodes and points
    Calculate the nodal mass $m_I$, momentum $\mathbf{p}_I$ and $\mathbf{v}_I$.
    Calculate the strain and spin rate $\dot{\boldsymbol{\varepsilon}}_p$, $\dot{\boldsymbol{\Omega}}_p$ of material points
    Calculate the stress $\boldsymbol{\sigma}_p$ of material points

(iii) **Coupling**: Update of material points
    Calculate the nodal force $\mathbf{f}_I^{ext}$, $\mathbf{f}_I^{int}$ and coupling force $\mathbf{f}_I^{cont}$
    Update the nodal momentum $\mathbf{p}_I$
    Update the position $\mathbf{x}_p$ and velocity $\mathbf{v}_p$ of material points

(iv) **Coupling**: Update of the SDEM particles
    Apply the $\mathbf{f}^{SDEM}$ to the SDEM particles
    Update the velocity, angular momentum
    Update the position of centre of mass for each SDEM particles

## 5. VALIDATIONS

A serial of tests is conducted in this section for the validation of MPM-SDEM method. They include three essential parts of the interaction between MPM and SDEM bodies: the conservation of energy, the contact force, and the granular-solid interaction. The first two tests are simulated using linear elasticity as the material property of the MPM; the last test is conducted with Drack-Prager model for the plastic deformation of granular material. The conservation of energy is crucial for the stability of the coupling. Contact forces between the MPM and SDEM bodies need to be correct since the contact handling method is unified under the DEM contact. It is also necessary to examine the plastic behaviour between the MPM and SDEM where the variation of contact relationships is far stronger than that of solid cases. Results are rigorously compared and analysed with analytical solutions. It helps to better understand the advantages and limitations of this method as a general scheme for the coupling between SDEM and MPM.

### 5.1 Conservation of energy

Two tests are designed to investigate the exchanges of momentum and the transferring between the kinetic and gravitational potential energy. The main purpose is to examine whether the energy of the system is increased by the collision between MPM and SDEM. The stability of the simulation can only be preserved if no extra energy is introduced into the system by the coupling algorithm. The other important purpose is to investigate if the coupling could well preserve the conservation of energy and how strong will the coupling affect the system energy. The system still has the numerical dissipation caused by pure DEM and MPM.

### 5.1.1 Exchange of momentum

The exchange of momentum is conducted by the 2D-elastic collision between MPM and SDEM disc. As illustrated in Figure 7 the MPM disc moving with a constant velocity of 2m/s towards the SDEM body, which is in a static state. The discs have the exact same shape, density, and size. The effect of gravity is eliminated from this test. Material properties and simulation parameters are listed in Table 1.

**Table 1** Parameters and material properties for the simulation of conservation of energy

| SDEM Parameters | | | | MPM (GIMP) Parameters | | | |
|---|---|---|---|---|---|---|---|
| $k_n$ | Normal stiffness | $6.0\times10^6$ | N/cm | $d_g$ | Grid interval | 0.3 | cm |
| $k_t$ | Tangential stiffness | $3.0\times10^5$ | N/cm | $r_p$ | Coupling radius | 0.1 | cm |
| $\mu$ | Frictional coefficient | 0.1 | | $n$ | Number of points | 5000 | |
| $\Delta t$ | Time interval | $2.0^{-4}$ | s | MPM material properties | | | |
| $V_d$ | Verlet distance | 0.2 | cm | $v$ | Poisson's ratio | 0.278 | |
| $a_i$ | Sphero radius | 2.0 | cm | **K** | **Bulk modulus** | $\mathbf{5.0\times10^2 / \times10^3}$ | **KPa** |
| SDEM material properties | | | | **G** | **Shear modulus** | $\mathbf{3.7\times10^2 / \times10^3}$ | **KPa** |
| $\rho_d$ | Density | 2.0 | g/cm² | $\rho_p$ | Density | 2.0 | g/cm² |

The MPM and SDEM share the same time interval and sequence; the coupling parameters for the collision handling are calculated with the same parameter of normal and tangential stiffness of SDEM. Two groups of material modulus, which are marked bold in Table 1, are used for the MPM disc. The bulk and shear modulus of the hard group is ten times larger than that of the soft case. This comparison is proposed to investigate the influence of the material properties to the conservation of energy. It could further demonstrate the ability for MPM-SDEM to simulate the soft-rigid multibody system.

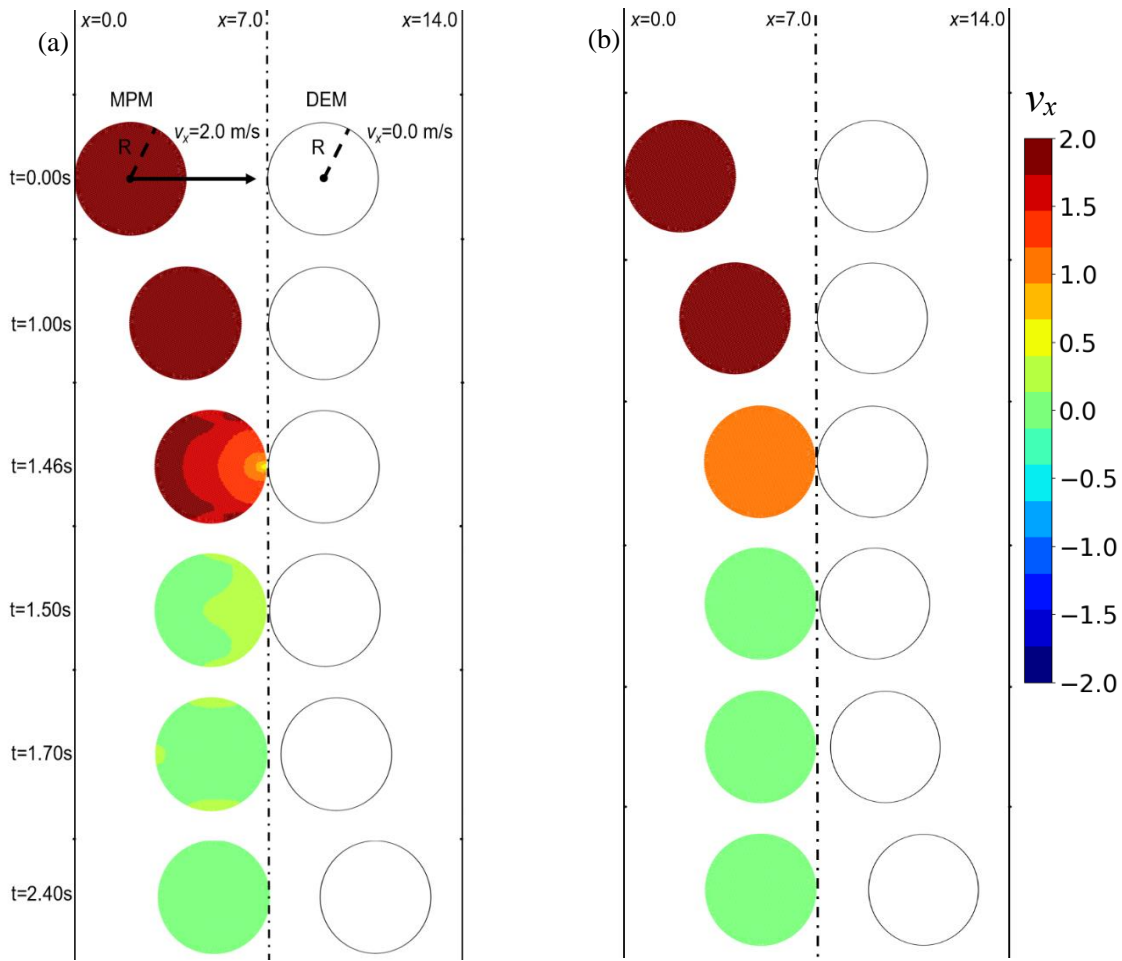

**FIGURE 7** The collision between the MPM and DEM disc at each time slice (Unit: cm/s). (a) the soft material case; (b) the hard material case. (color encodes velocity)

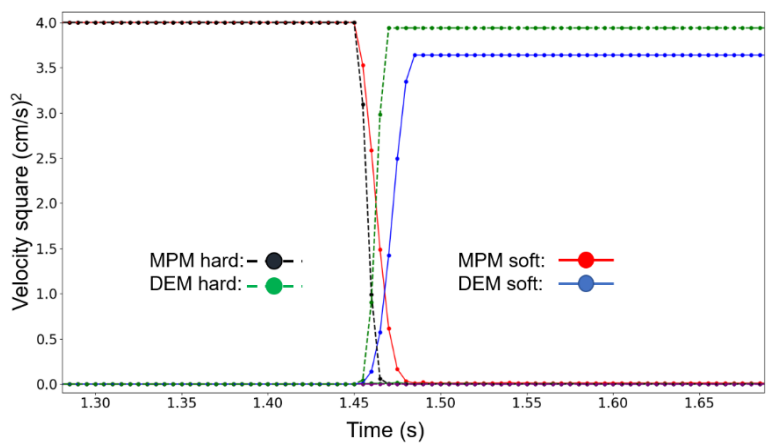

**FIGURE 8** The variation of velocity square of DEM and MPM disc with different material properties.

It can be observed from Figure 7 (a) that the contact between two bodies is captured by the coupling algorithm and the collision is calculated with the contact force. The velocity of the MPM disc drops rapidly during the collision and transfer its kinetic energy to the DEM disc at $t$=1.46s and 1.50s. Part of the energy is stored in the MPM body in a form of strain energy because some of the material points at $t$=1.70s still have a small velocity. This part of the energy is dissipated in the MPM disc since the collision is completed. It dissipated due to the *P-G* transferring scheme and eventually disappear(e.g. $t$=2.4s). This phenomenon is consistent with the variation of squared $v_x$ in Figure 8, where kinetic energy is largely transferred into the DEM disc, which obtained a $v_x$=1.92m/s. It still shows a clear loss of kinetic energy as these part are not transferred to DEM after the collision.

For the hard material case in Figure 8, the conservation of energy is much better than that of the soft case. The collision happened in a short time of duration as it is shown in Figure 7 (b) after $t$=1.46s. The velocity of material points becomes zero after the collision. The SDEM disc obtained a $v_x$=1.992m/s. The kinetic energy is much closer to the analytical solution of the perfect elastic collision.

**5.1.2    Transferring between the gravitational potential and kinetic energy**

The test for the potential-kinetic energy transferring is conducted by dropping an MPM elastic disc at the spheropolygon boundary. The spheropolygon element is set as an unmoveable elastic boundary. It can be seen from Figure 9 that the MPM disc hits the boundary and bounces back at the spheropolygon boundary. The contact is detected by the algorithm and the collision are properly calculated. The simulation parameters and material properties are the same as Table 1; the gravitational acceleration is -10.0 cm/s$^2$.

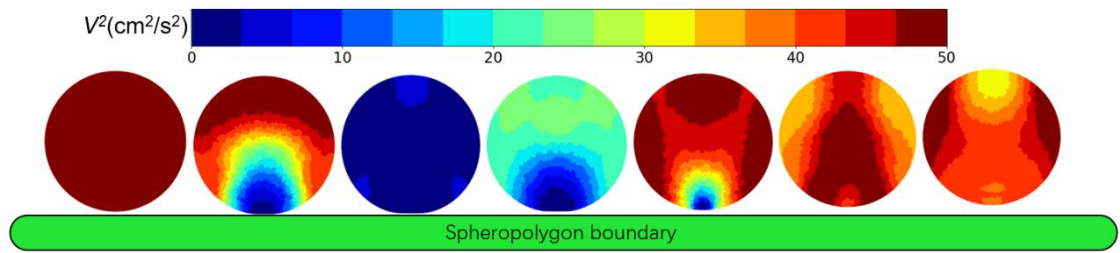

**FIGURE 9** The soft elastic disc dropping at and bouncing back by the spheropolygon boundary

The variations of the vertical velocity square are presented in Figure 10. The hard disc is shifted with +0.5s on the time axis for the comparison. Figure 10 indicates that the soft discs still have a clear loss of energy after the collision. The hard disc case, similarly, has a better performance for the conservation of energy; less energy is dissipated during the collision; its velocity almost reaches the theoretical limit. It can also be observed from Figure 11 that soft disc experienced a longer time to complete the transferring between kinetic and potential energy. For both soft and hard disc, no extra energy is generated by the collision since the squared velocity after the collision is lower than the analytical limit. It indicates the coupling algorithm can maintain its stability. This results further supports the fact that the material property has a certain influence on the conservation of energy. The low value of elastic modulus could generate a strong loss of energy after the coupled collision. It may cause a problem for the simulation that requires a high velocity (e.g. impact engineering) but could still be well-applied for the simulation of quasi-static or low-velocity issues.

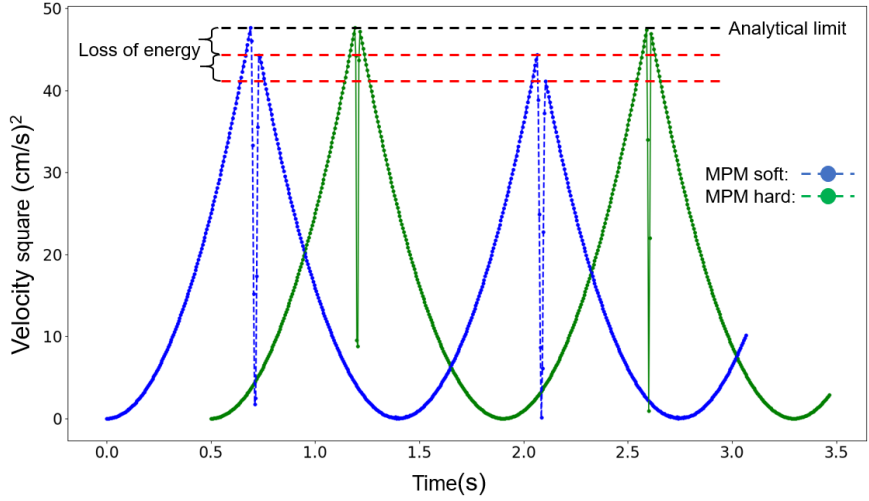

**FIGURE 10** The change of velocity squared of the MPM disc; the hard disc is shifted with a 0.5s on the time axis for the comparison.

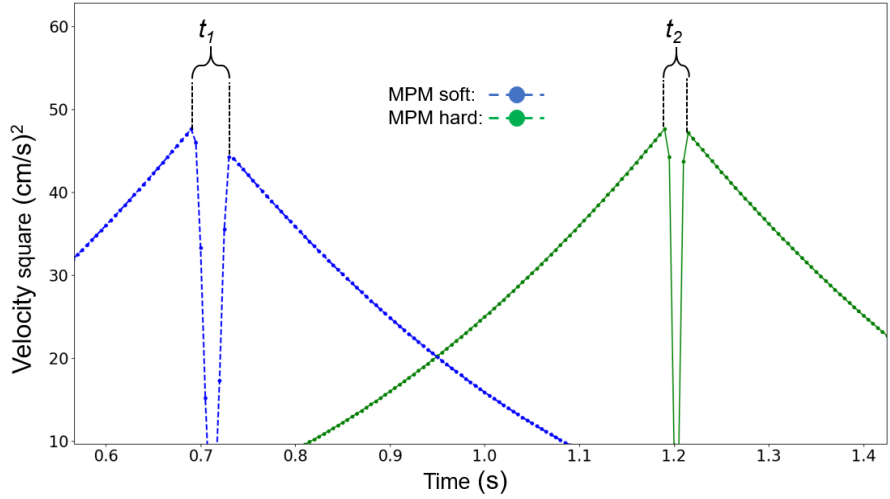

**FIGURE 11** The time gap $t_1$ and $t_2$ of collision obtained with soft and hard MPM disc case.

## 5.2 Coupling force

The coupling of the MPM and SDEM is conducted through the contact force that calculated based on the contact model of DEM. It distinguishes the coupling method developed in this paper from the existing method, which can be used for the quasi-static and dynamic cases. The value of normal and frictional coupling force between the MPM and SDEM are examined in this section. Especially, the contact force's dependence on the grid size is investigated here with four different values of grid interval. Simulation parameters and material properties are given in Table 2; integer $i$ denotes the variation of the grid size, which ranges from 0 to 3 and marked in bold.

Table 2 Parameters and material properties for the simulation of MPM-SDEM contact force

| SDEM Parameters | | | | MPM (GIMP-PIC) Parameters | | | |
|---|---|---|---|---|---|---|---|
| $k_n$ | Normal stiffness | $6.0\times10^6$ | N/cm | $d_g$ | **Grid interval** | **0.35+0.05×$i$** | cm |
| $k_t$ | Tangential stiffness | $3.0\times10^5$ | N/cm | $r_p$ | Coupling radius | 0.1 | cm |
| $\mu$ | Frictional coefficient | 0.1 | | $n$ | Number of points | 2601 | |
| $\Delta t$ | Time interval | $2.0^{-4}$ | s | | MPM material properties | | |
| $V_d$ | Verlet distance | 0.2 | cm | $v$ | Poisson's ratio | 0.2558 | |
| $a_i$ | Sphero radius | 0.5 | cm | $K$ | Bulk modulus | $6.0\times10^3$ | KPa |
| General Parameter | | | | $G$ | Shear modulus | $3.5\times10^3$ | KPa |
| $g$ | Gravitational acceleration | 100.0 | cm/s² | $\rho_p$ | Density | 2.5 | g/cm² |

### 5.2.1 Normal force

The normal contact force is examined by placing an MPM elastic square on the spheropolygon boundary. Reaction forces are generated due to the gravitational force and, based on the coupling algorithm, applied on each material point at the bottom of the square. Figure 12 indicates that the contact relations between material points and the boundary are properly captured; the contact force is equally applied at the points that are in contact with the spheropolygon boundary. There are in total 51 material points near the boundary and each one has a 0.4902N normal force as the reaction force for the gravity.

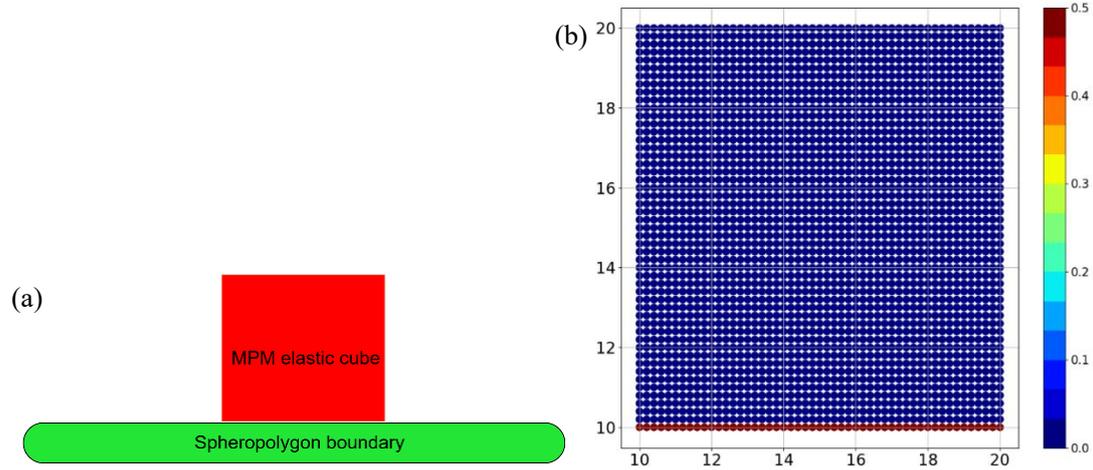

**FIGURE 12** Normal forces generated between the elastic MPM block and spheropolygon boundary (Unit: N); $d_g$=0.35 and $t$=3.5s. The normal forces are generated at the bottom of the MPM elastic square where the material points are in contact with the spheropolygon boundary; the magnitude of the force generated each point are all equal to 0.4902N (red value in colour bar); the other material points have zero force value (blue value in colour bar) since they are not in contact with the boundary.

The total reaction force for different grid size is shown in Figure 13. It indicates that the simulated value of normal force has a strong vibration at the beginning of the simulation. Such vibration is caused by the initial zero-overlapping configuration between the boundary and the elastic block. The normal force gradually converges to the analytical solution. The progress indicates that the static equilibrium is searched by the contact algorithm and eventually reached. The kinetic energy is dissipated due to the viscoelastic contact model of DEM and the numerical damping within the point-grid transferring scheme of MPM. The energy dissipation here is a positive factor, which helps the system reaches a static state and provides the correct force information. This tendency can be observed for all four cases with different grid sizes. The larger size of the grid only improves the intensity of the energy dissipation but does not change the convergence of the contact force; the correctness of the contact force is independent of the size of the background grid. The MPM and SDEM can be properly coupled through the DEM contact scheme.

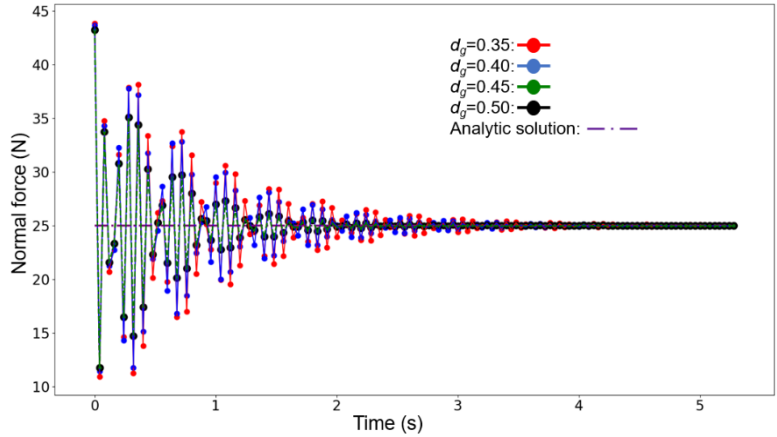

**FIGURE 13** The total normal force for different background grid size.

### 5.2.2 Frictional force

The same model is used for simulating the frictional force. A constant velocity of $V_x$=2.0m/s is applied to the same MPM square. It is sliding towards the positive direction of the *x*-axis and generates a friction force. Its friction coefficient between the MPM and DEM boundary are set as $\mu$=0.3. It can be observed from Figure 14 that the friction force applies to the material points located at the bottom of the elastic MPM square. The magnitude of the friction force is $\mu F_n$=-7.5N, which equals to -0.14706N for each material points that in contact with the boundary. This result is consistent with the result of the normal-force test.

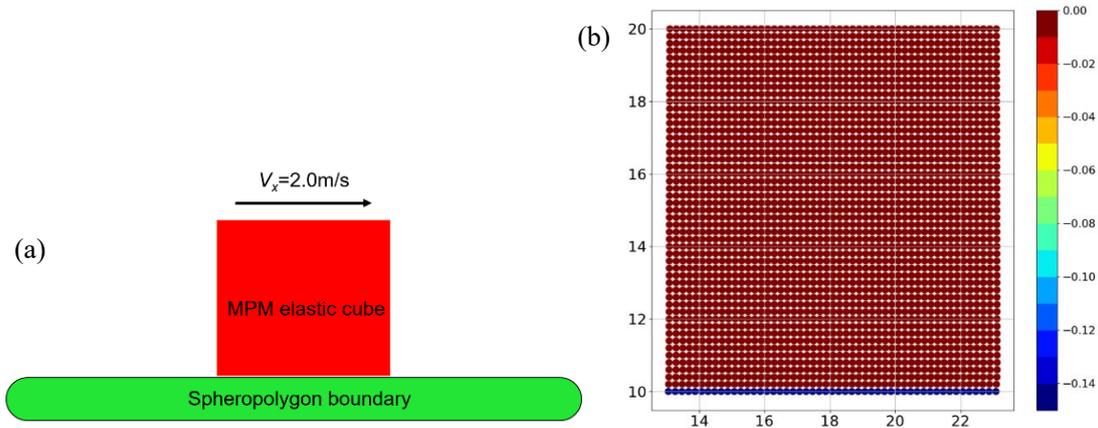

**FIGURE 14** Frictional forces generated between the elastic MPM block and spheropolygon boundary (Unit: N); $d_g$=0.35 and *t*=3.5s. The frictional forces are generated at the bottom of the MPM elastic square where the material points are in contact with the spheropolygon boundary while sliding; the magnitude of the frictional force on each point are all equal (blue value in colour bar) since their relative position to the boundary is the same; the other material points have zero frictional force (red value in colour bar) since they are not in contact with the boundary.

The variation of total frictional force is shown in Figure 15. The test for the effect of grid size has been clarified in the normal force test and therefore not repeated here. The grid size $d_g$=0.50 is used to quickly reduce the vibrations generated by initial configuration. Similarly, the friction force also converges to the analytical value with the progress of the simulation. The vibration energy is dissipated and the MPM square reaches a quasistatic state; both normal and friction force are correctly calculated.

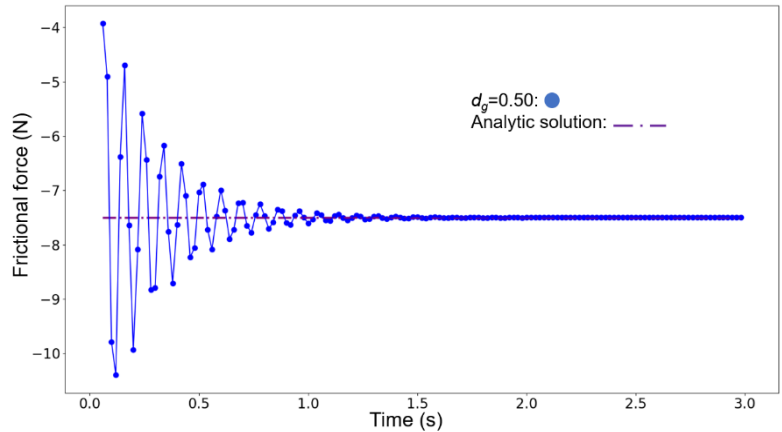

**FIGURE 15** The total frictional force for $d_g$=0.5

## 5.3 Granular flow

The coupling performance between the SDEM and the MPM granular material are tested with the silo flow model. The Drucker-Prager model is used to simulate the plastic behaviour of the non-cohesive dry sand which represented with MPM. The SDEM are used as the silo and container of the granular flow. Five different diameters of the silo neck are used to compare the results with the 2D Beverloo law. The main reason for conducting these tests is because the contact relationship for the plastic MPM-SDEM case is far more changeable than that of elastic MPM-SDEM cases. Contacts are constantly generated and deleted in the granular flow. Therefore, it is necessary to test whether coupling simulation is reliable. Material points, although representing a continuous area, are used to approximate the assemble of discrete bodies (i.e. sand or rock pile). Three different number of material points are used to investigate its influence on mass transportation. The simulation parameters and material properties are given in Table 3.

Table 3 Parameters and material properties for the simulation of MPM-SDEM granular flow

| | SDEM Parameters | | | | MPM (APIC) Parameters | | |
|---|---|---|---|---|---|---|---|
| $k_n$ | Normal stiffness | $6.0 \times 10^6$ | N/cm | $d_g$ | Grid interval | 0.35 | cm |
| $k_t$ | Tangential stiffness | $3.0 \times 10^5$ | N/cm | $r_p$ | Coupling radius | 0.1 | cm |
| $\mu$ | Frictional coefficient | 0.2 | | $n$ | Number of points | $8/9/10 \times 10^3$ | |
| $\Delta t$ | Time interval | $2.0^{-4}$ | s | | MPM material properties | | |
| $V_d$ | Verlet distance | 0.2 | cm | $v$ | Poisson's ratio | 0.2358 | |
| $r_d$ | Sphero radius | 0.5 | cm | $K$ | Bulk modulus | $5.0 \times 10^2$ | KPa |
| | General Parameter | | | $G$ | Shear modulus | $3.2 \times 10^2$ | KPa |
| $g$ | Gravitational acceleration | 10.0 | cm/s$^2$ | $\rho_m$ | Density | 1.5 | g/cm$^2$ |
| | | | Drucker-Prager model | | | | |
| $\sigma_t$ | Tensile strength | 0.0 | MPa | $\phi$ | Friction angle | 35.0 | degree |
| $\psi$ | Dilation angle | 25.0 | degree | | | | |

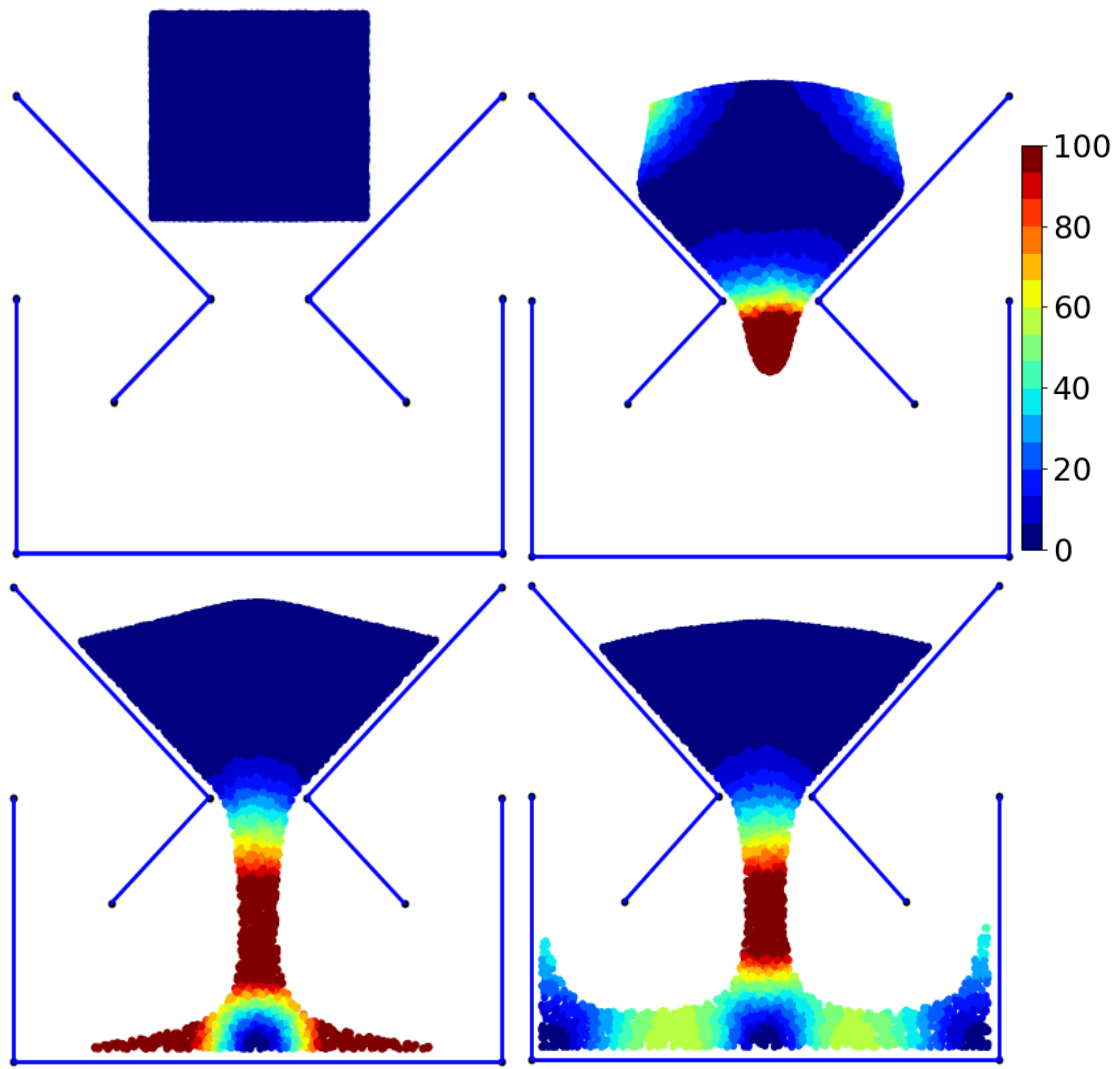

**FIGURE 14** The $v^2$ for the dry sand silo flow at $D_0$=0.04m (Unit: cm$^2$/s$^2$)(a) $t$=0.0s (b) $t$=1.0s (c)$t$=2.0s (d)$t$=3.0s

As illustrated in Figure 14, the MPM dry sand flows through the bottleneck of the silo and reaches a steady flowing state; the sand drops at the bottom and held by the SDEM container. It indicates that the coupling algorithm remains stable for the granular material. The contact can be effectively detected and properly handled.

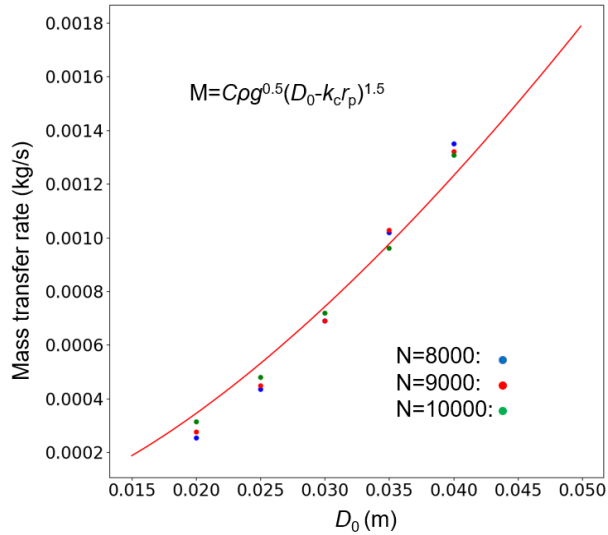

**FIGURE 15** The mass transfer rate for different silo neck diameter and number of material points; $C=0.58$ and $k_c=2.2$ are the empirical coefficient; $D_0$ is the diameter of the silo neck.

The mass transfer rates for different sizes of silo neck are shown in Figure 15. Simulations with different numbers of material points are performed for each diameter. Results indicate that the simulated mass transfer rates agree well with the analytical solution of the 2D Beverloo law[43]. The rate increases with the increase of the diameter in a 3/2 power law relation. The number of material points does not have a strong influence on the results since MPM is a continuous numerical method. Although there is still a deficiency for using the Drucker-Prager model as indicated earlier, the coupling method is reliable for simulating granular-solid interaction.

### 5.4 Discussion

The studies in this section prove that the coupling between MPM and SDEM can be properly calculated under the contact scheme of DEM. Results agree well with the analytical solutions for tests of conservation of energy, contact force, and granular flow. The conservation of energy is affected by the material property but properly maintained in general. No extra energy is generated from the collision between material points and SDEM contact layer, which indicates the stability of the method. Contact forces, as the communicating value for the coupling, are correctly provided. Such correctness guarantees the proper interaction between a continuous (MPM) and a discontinuous (SDEM) method. Tests of granular flow indicate that the coupling could effectively detect and handle the highly changeable contact relations between the granular media and SDEM particles. Therefore, MPM-SDEM coupling method can be applied for many numerical simulations, which have a large difference in size or Young's modulus.

### 6. APPLICATIONS OF MPM-SDEM

Two applications of MPM-SDEM coupling method are provided in this section to further demonstrate its ability to simulating the problems that regarded as challenging for traditional methods. In the first application, we simulated the motion of wooden blocks under the impact of granular materials. The simulation is conducted with the same parameters and configurations used by Liu et al[24] for the coupling of MPM and DEM. It is a meaningful comparison to study the difference between two coupling methods. The second application is a quasi-static compression of a fine-particle granular matrix. It is a typical structure for Geomechanics that has exhibited complication mechanical features and poorly studied due to the limitation of numerical and experimental tools.

**6.1 Granular flow and motion of blocks**

The motion of wooden blocks under the impact of the granular flow is simulated in two dimensions. This test was originally conducted by Liu et al[24] for the validation of a coupled MPM-DEM method. In their method, each DEM block is represented by 9 visual material points to couple its interaction with MPM granular flow based on the exchange of momentum. The configurations of the test are given in Figure 16. The granular flow will be released and impact three piled-up wooden blocks. For the granular material, the gravitational potential is transferred to the kinetic energy and eventually exerts on the blocks through the point-grid projection on the mutual background grids. Two upper blocks will be pushed to the right side and obtain angular momentum; Block.3 is glued to the ground and thus unmovable. The variation of the inclining angle of block No.2 is recorded and compared with experimental data.

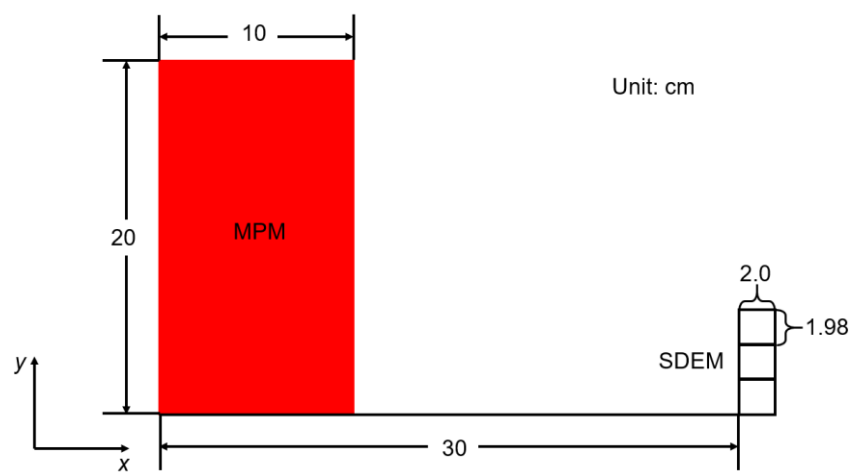

**FIGURE 16** The initial configuration of the granular flow and SDEM blocks

In this section, this test is reconducted using our MPM-SDEM coupling scheme. The results are compared and rigorously analysed with both numerical and experimental data from the existing. It will further illustrate the advantages of using the contact force as the coupling connection. The material properties and simulation parameters are given in Table.3 with a few modifications for the simulation: $g$=10m/s, $\psi$=0°, $n$=8000 and $\phi$=22.0°.

The movements of granular flow and blocks for the simulation are shown in Figure 17. The time for each screenshot is taken at the exact same time of Liu et al's study to provide a clear visual comparison. They indicate that both the collapse of the granular pile and the motion of the blocks agree well with the existing numerical and the experimental data at each compared time step. The sand flow reaches the pile of blocks at $t$=2.5s; the blocks start to move and rotate due to the impact.

The block No.2 touches the ground at $t$=4.0s and block No.1 touches the ground after $t$=4.5s.

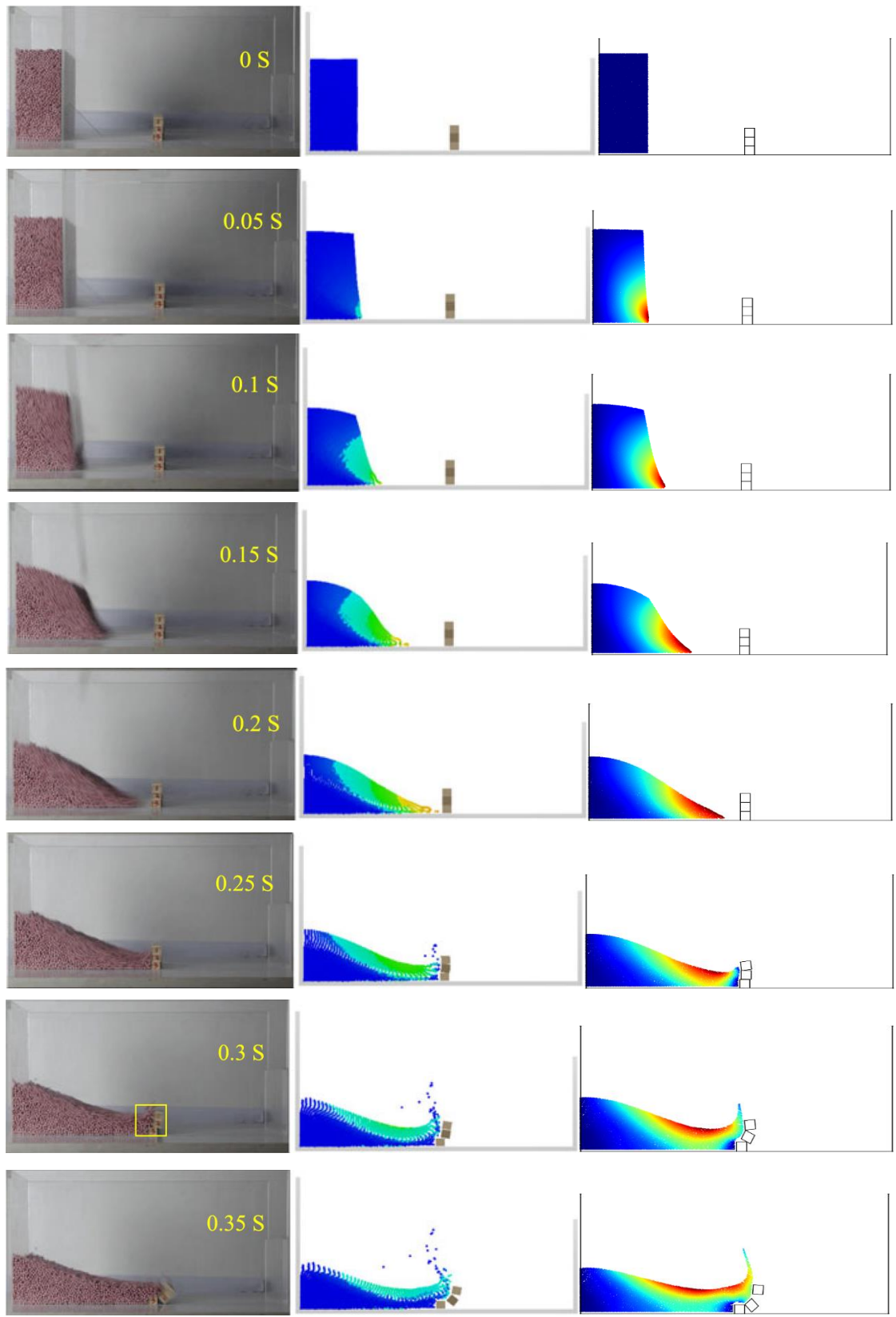

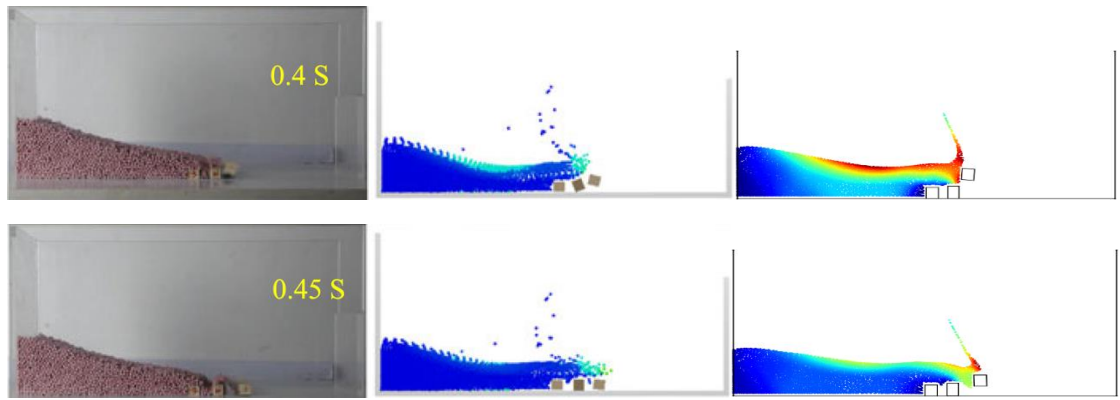

**FIGURE 17** Comparison among the Liu et al's experiment(left) and numerical results(middle), and MPM-SDEM (right) simulation of granular flow impacting blocks at different time steps.

The rotation of block No.2 is recorded and compared with the numerical and experimental data in Figure 18. It can be observed that comparing with the existing method inclination angles of MPM-SDEM are closer to the experimental results at each measured time step, except $t$=0.3s. Such better performance is particularly obvious at $t$=0.25 and $t$=0.40s. The differences between MPM-SDEM and experimental data are smaller than 4.0°, where the different almost reaches 20.0° for the existing method. Furthermore, the variation curve of the experiment is smoother than that of Liu's method. It indicates that the actual rotation and changes of angular momentum of block No.2 are relatively continuous. The result of the existing coupling method exhibits jumps of the angular velocity. This discontinuity is particularly strong between $t$=0.2s to 0.25s.

As discussed earlier, this problem is caused by two aspects. First, the DEM blocks are insufficiently represented with only 9 material points. There is a strong loss of accuracy when calculating the momentum transferred from granular flow to the highly simplified DEM block. Secondly, the coupling method has a strong dependency on the background grid. The angular momentum will be highly compromised if the gird size is too large; however, if the grid size is too small the material point at the centre of mass can be ignored and receives zero momentum from the granular flow. The grid size has a strong influence on the simulation, yet the optimal value is very hard to be determined; neither increase nor decrease the grid size can guarantee the increase of accuracy. The improvement can be made by increasing the number of material points to represent the DEM block. But that strategy makes a large compromise of computational efficiency; both detection and calculation of the contact become far more expensive. Therefore, this method is only reliable for certain cases.

The variation curve provided by MPM-SDEM shows a continuously changing pattern like the experimental data. No sudden jumps of angular velocity for the SDEM block. Because the coupling is conducted through the contact force. Neither contact detection nor force calculation has a strong dependency on the grid. The angular momentum is better preserved during the impact process since the transferring of the momentum is not limited by the grid. Material points of the granular flow can transfer its kinetic energy to the edges of blocks at any position if their Euclidian distance is recognized as in contact. Contact detection and force can be obtained using solely the Euclidian distance. MPM-SDEM shows a superior performance comparing with the existing method and should be reliable for a variety of application due to its unique advantages.

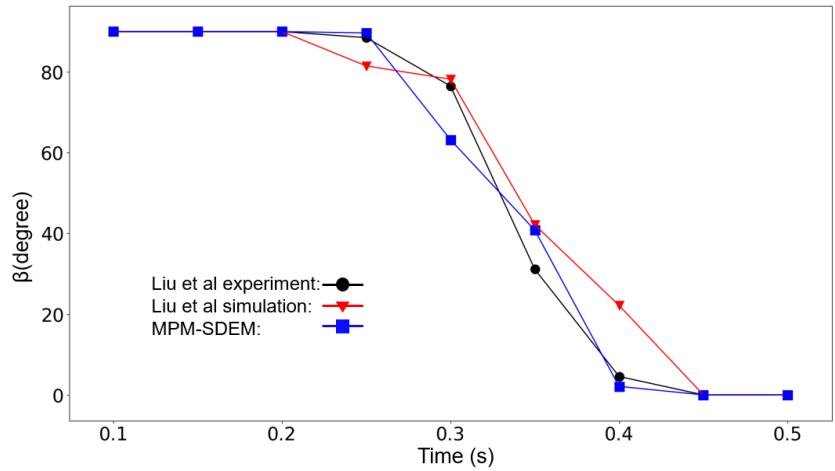

**FIGURE 18** The temporal variation of block No.2's inclining angle *β*.

**6.2 Uniaxial compression of the Granular matrix**

An important application for the MPM-SDEM coupling method is the simulation for granular matrix. It is consists of particles in very different sizes and thus produces different physical properties as the assembling of them. The size ratio between large particle and fines could reach $10^{5-6}$ level. It is highly computational expensive to simulate this structure with pure DEM method. Pure continuous methods also have difficulty to properly simulate the interaction between large particles and fines. Studies[1, 44] have been done with DEM in 2D and certain progress has been made based on the results. However, both particles and the fines are poorly represented with a limited number and therefore compromises the accuracy of the results.

The MPM-SDEM method provides a better way to simulate the micro-mechanism of the granular matrix. It is a scale-crossing method that could represent the particles in an extreme size ratio. Large particles are represented by SDEM, the irregular shape can be properly considered. Fines, which exists in the voids between the particles, are simulated with MPM. The mechanical properties of the fines are governed by the *D-P* model. Its effects no longer need to be represented with a large amount of the tiny particles while the discrete interaction of large particles can still be preserved. An example of the uniaxial compression test is conducted to investigate the effect of fine percentage in the granular matrix. Large particles are generated with irregular pentagons. The parameters are the same as Table 4 except the density of SDEM particles is changed to 2.2 g/cm$^2$ .

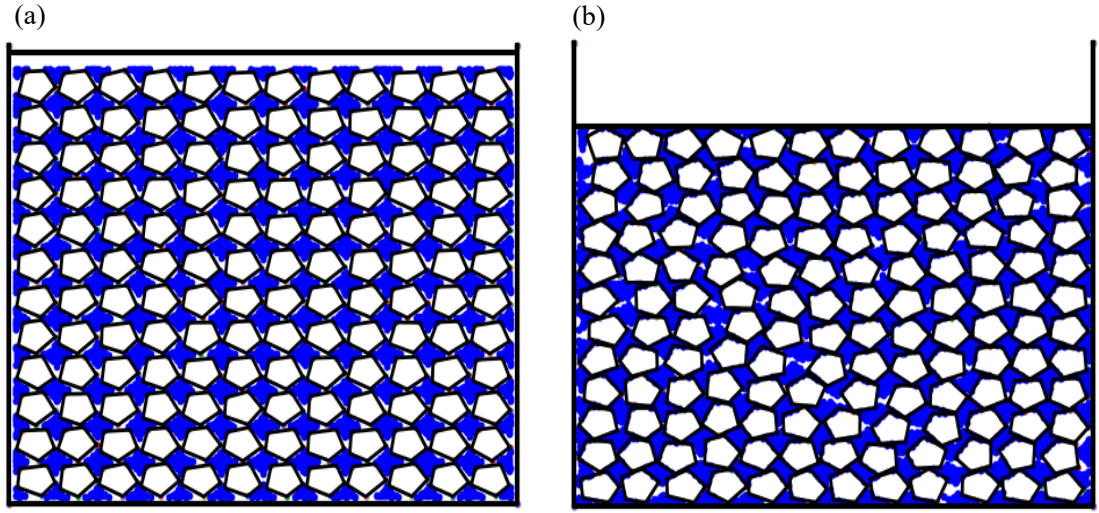

**FIGURE** 19 The uniaxial compression of granular matrix consist of fines and particles. The container size of 61×61 cm; the compression bar is moving with a constant velocity of $v_y$=-0.5cm/s. (a) The initial configuration of the granular matrix (b) granular matrix at $t$=22.0s

The basic structure of the granular matrix sample is shown in Figure 19 SDEM particles are the skeleton while MPM fines exist in the voids among them. Both of the components bear the compressive load from the top. At the early stage, the compression mainly exerts on the SDEM particles since the void space is not fully saturated. SDEM particles will start to slowly move and rotate. It can be regarded as a rearrangement process for the skeleton part of the granular matrix. The void space is shrinking due to this rearrangement. MPM fines start to carry the compressive load by interacting with SDEM particles if the void space is small enough. This phenomenon is one of the reasons that the granular matrix exhibits a relatively complex mechanical behaviour.

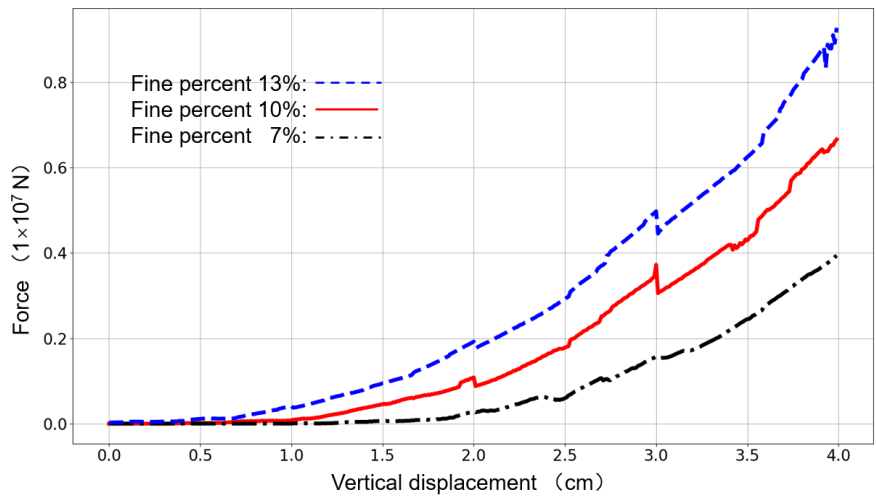

**FIGURE** 20 The force-settlement curve for different percentage of fine content.

The compression force-settlement curves are shown in Figure 20. It can be observed from the results that the fine percentage has a major impact on the compressibility of the granular matrix, which would further affect the result for measuring Young's modulus. The basic skeleton structure of the SDEM is preserved because the void space is not saturated by fines and the skeleton structure thus should not be strongly affected by a small change of fine content. It indicates that the fine content could effectively change the mechanical property; a higher fine percentage largely reduces the compressibility of the granular matrix and, therefore, exhibits a higher Young's modulus at the macroscale. The test provided a good demonstration of the potential application of MPM-SDEM method. More rigorous numerical tests with comprehensive consideration and higher accuracy will be conducted in the future study.

## 7. CONCLUSIONS

An advanced MPM-SDEM coupling method is developed for simulating the multi-body system with a large ratio of size and material modulus. Especially the interaction between solid and granular materials. This continuous-discontinuous algorithm has advantages of the MPM for dealing with large deformation and the efficiency of SDEM for handling the movement of the irregular near-grid body. This method is conducted through the contact force between SDEM particles and material points. Both contact detection and the force calculation are unified under the contact scheme of the SDEM method. A material point is identified as a circular discrete element (IMP) if its Elucidian distance to the centre of mass of an SDEM particle is smaller than the Verlet distance. The normal and tangential forces of the contact can then be calculated based on the intersection and relative displacement between the SDEM particle and the IMP. For MPM simulation, this contact force will be applied to the IMP as an external boundary force term. The movement of material points is influenced and constrained by the contact forces, which is also added to the total force vector on the corresponding SDEM particle to determine its velocity, rotation, and position.

A serial of validation tests is conducted with two different constitutive models, linear elasticity and Drucker-Prager model. The conservation of energy, contact force, mass transfer rate in granular flow is investigated and analysed with analytical solutions. Results indicate that the conservation of energy is well preserved by the coupling method through the contact force. However, a certain amount of energy will lose during the collision if the material modulus of MPM is extremely low. Both normal and friction force between the MPM and SDEM particle are correctly calculated after the vibration generated by the initial configuration. Such vibration energy can be effectively dissipated with a larger size of the background grid and transferring format. Granular flow simulated by MPM also provides a good agreement with the Beverloo law for the mass transfer rate.

Two representative applications are presented. The motion of wooden blocks under the impact of the granular flow is conducted and compared with an existing MPM-DEM coupling method and experimental data. The variation of inclination angle provided by MPM-SDEM is more continuous and shows a better agreement with the experimental data. It overcomes the problems of the grid dependency in the existing MPM-DEM algorithm. Uniaxial compression tests for a granular matrix are performed. It further demonstrates that MPM-SDEM coupling method provides a scale-crossing solution for simulating the complex mechanical behaviour of the granular matrix. Both continuous behaviour of fines and discrete features of large particles can be well preserved in the simulation with optimal efficiency. The results indicate that the fine percentage in a granular matrix has a major influence on its macroscale mechanical properties. In conclusion, the MPM-SDEM has its unique advantage for simulating the multi-body interaction with highly different size and material modulus. We anticipate it would make a good contribution to the study of geomechanics and CG production.

**APPENDIX**

The three-dimensional weighting functions for the GIMP is written as:

$$S_{Ip} = S_{xIp}(\mathbf{x}_p) \cdot S_{yIp}(\mathbf{x}_p) \cdot S_{zIp}(\mathbf{x}_p), \tag{A1}$$

where $S_{iIp}$, which is calculated based on the constant characteristic function and linear shape function, is defined as:

$$S_{iIp} = \begin{cases} 0, & x_{ip} - x_{iI} \leq -(L+l_p) \\ [L+l_p+(x_{ip}-x_{iI})]^2/(4Ll_p), & -(L+l_p) < x_{ip} - x_{iI} \leq -L+l_p \\ 1+(x_{ip}-x_{iI})/L, & -L+l_p < x_{ip} - x_{iI} \leq -l_p \\ 1-[(x_{ip}-x_{iI})^2+l_p^2]/(2Ll_p), & -l_p < x_{ip} - x_{iI} \leq l_p \\ 1+(x_{ip}-x_{iI})/L, & l_p < x_{ip} - x_{iI} \leq L-l_p \\ [L+l_p-(x_{ip}-x_{iI})]^2/(4Ll_p), & L-l_p < x_{ip} - x_{iI} \leq L+l_p \\ 0, & x_{ip} - x_{iI} > L+l_p \end{cases} \tag{A2}$$

where $L$ is the spacing of the background grid and $l_p$ is the half size of the square influential area defined by the characteristic function $\chi_p$. This weighting function is used for all the simulations of solid material.

Since the characteristic function is Dirac delta function $\chi_p(\mathbf{x})=\delta(\mathbf{x}_p)$, the three-dimensional weighting function for the APIC format is also the shape function which is using the quadratic kernel:

$$N_{Ip} = N_{xIp}(\mathbf{x}_p) \cdot N_{yIp}(\mathbf{x}_p) \cdot N_{zIp}(\mathbf{x}_p), \tag{A3}$$

where $N_{iIp}$ is written as:

$$N_{iIp} = \begin{cases} (1/2)(|x_{ip}-x_{Ip}|/L)^3 - (|x_{ip}-x_{Ip}|/L)^2 + (2/3), & 0 \leq |x_{ip}-x_{Ip}|/L < 1 \\ (1/6)(2-|x_{ip}-x_{Ip}|/L)^3, & 1 \leq |x_{ip}-x_{Ip}|/L < 2 \\ 0, & 2 \leq |x_{ip}-x_{Ip}|/L \end{cases} \tag{A4}$$